\documentclass[final,5p,times]{elsarticle}

\usepackage{epsfig}
\usepackage{xcolor}
\usepackage{url}

\usepackage{amssymb}
\usepackage{amsmath}

\def\ket#1{  \left\vert  #1   \right\rangle   }
\def\bra#1{  \left\langle  #1   \right\vert   }

\journal{Results in Physics}

\begin{document}

\begin{frontmatter}

\title{Bell Test and Quantum State-Space Geometry}

\author{Michael Siomau} 

\affiliation[1]{organization={Rabdan Academy},
            city={Abu Dhabi}, 
            country={United Arab Emirates}}
\affiliation[2]{organization={Institute for Theoretical Physics},
            city={ETH Zurich}, 
            country={Switzerland}}

\begin{abstract}
We explore a minimal phenomenological model of coupling between the geometry of the classical spacetime and the quantum state space. Assuming a simple structure of the combined manifold, bilinear interaction between the corresponding geometric scalars and the simplest case of the two-qubit entanglement dynamics, we derive an illustrative upper bound on the geometric coupling constant from the requirement of the Clauser-Horne-Shimony-Holt (CHSH) inequality violation over a finite time interval.

\end{abstract}




\end{frontmatter}

There is no comprehensive theory of quantum gravity that consistently unifies quantum mechanics with general relativity in all relevant physical regimes \cite{Kiefer}. The state of a quantum system defined in an abstract Hilbert space is not a local observable in the same sense as position, momentum, or a field, which makes it impossible to use the (classical) spacetime metric $g_{\mu\nu}(x)$ on a four-dimensional Lorentz manifold to construct a quantum metric \cite{Witten}. Furthermore, the unitary evolution of quantum states suggests considering whether the corresponding spacetime geometry should incorporate an appropriate complex structure. Provost and Vallee were the first to explore a complex metric in the space of quantum  parameters \cite{Provost}. They showed that its real part corresponds to a gauge-invariant Riemann metric, while its imaginary part has a symplectic structure, now known as the Berry curvature \cite{Berry}. Shortly, a closely related Fubini-Study metric, defined on the projective Hilbert space, was used to characterize the geometric phase \cite{Anandan}. These metrics are commonly used as the starting point for studying the geometry of quantum state manifolds and are sometimes called the complete quantum metric \cite{Hetenyi}. 

But the metric itself does not specify quantum measurement theory, i.e. how to extract probabilities from the quantum state dynamics. This motivates a deep search for the connection between the geometric properties of spacetime with quantum information measures \cite{Neukart} and decoherence \cite{Anast,Oppenheim}. However, the formulation of a general theory of measurement and observables in quantum gravity remains an open problem \cite{Kiefer}. 

The theory of quantum gravity should incorporate both classical and quantum metrics, reproducing the corresponding predictions in the limiting regimes. The simplest assumption is that the unifying metric is diagonal

\begin{equation}
 \label{metric}
\mathcal{G}
=
\begin{pmatrix}
g_{\mu\nu} & 0\\
0 & h_{ab}
\end{pmatrix}
\end{equation}
where \(g_{\mu\nu}\) is the spacetime metric on a four-dimensional Lorentz manifold \(\mathcal{M}\), and \(h_{ab}\) is defined on a quantum state-space Hilbert manifold \(\mathcal{H}\). Although this hypothesis may seem naive, (\ref{metric}) does not violate the existing criterion for  metrics to be physical \cite{Kontsevich}. 

From Eq.~(\ref{metric}) alone it follows that the total geometric action \cite{Kiefer} should have an interaction term with some coupling constant, fundamentally restricted by quantum dynamics and quantum non-locality. In this brief report, we illustrate the bound on this constant. We introduce a bilinear interaction term in the geometric action by coupling the Ricci scalar to the Berry curvature. Identifying the latter with a concurrence-based entanglement functional for pure two-qubit states, we model the effect of the coupling as a curvature-dependent correction to the effective decoherence rate of a Werner-type channel. The requirement that the CHSH value remain above the classical bound on a Bell-test timescale yields an explicit upper bound on the coupling constant. Finally, we briefly examine how the interaction term can be interpreted as an entanglement correction to the effective spacetime curvature. 

From the definition of the metric (\ref{metric}), it follows that the total geometric action can be written as
\begin{equation}
S = S_g + S_q + S_{\mathrm{int}} .
\label{action}
\end{equation}

The first term is the Einstein--Hilbert action.
\begin{equation}
S_g = \frac{1}{16\pi G}\int_\mathcal{M} dx\,\sqrt{-g}\,(R-2\Lambda),
\label{EHaction}
\end{equation}
where $g=\det\{g_{\mu\nu}\}$, $R$ is the Ricci scalar curvature, $G$ is Newton's gravitational
constant and $\Lambda$ is the cosmological constant, while $c\equiv1$ and $\hbar\equiv1$.

The second term is a geometric action analogous to 
(\ref{EHaction}),
\begin{equation}
S_q = \eta \int_{\mathcal{H}} dx\,\sqrt{-h}\,B,
\end{equation}
where $h=\det\{h_{ab}\}$, $B$ is the Berry curvature of the quantum geometric tensor, and $\eta$ is a constant.

Without an explicit map between spacetime and quantum-state points, the interaction integral $S_{\mathrm{int}}$ over the manifold $\mathcal{M}\otimes\mathcal{H}$ is not well defined. Instead, we employ the simplest local, coordinate-invariant alternative, i.e. 

\begin{equation}
S_{\rm int} = \lambda \int_\mathcal{M} dx\sqrt{-g}\,R\;\bar{B},
\end{equation}
where
\begin{equation}
\bar{B} \equiv \int_{\mathcal{H}} dx\,\sqrt{-h}\,B.
\label{interaction}
\end{equation}
Here $\lambda$ sets the strength of the coupling between the curvatures. $\bar{B}$ plays the role of a state-space-averaged entanglement content that the spacetime curvature $R$ couples to.

While the Ricci scalar curvature $R$ can be defined in the standard way through the Riemann curvature tensor and the Ricci tensor, for the Berry curvature $B$ we use its connection with entanglement through a geometric quantum-resource functional $B\equiv E[\psi]$, which takes an especially simple form for pure two-qubit states \cite{Levay}, i.e.

\begin{equation}
E[\psi] = \kappa C^2,
\end{equation}
where $C$ is the concurrence \cite{Wootters} and $\kappa>0$ is a normalization constant. We would like to stress that this assumption is crucial for evaluating (\ref{interaction}). 

To proceed, we need to define a decoherence model. The
fact that we replaced $B$ with $\bar{B}$ in the definition of the interaction integral (\ref{interaction}), allows us to use the Lindblad master equation \cite{Manzano} approach with a modified curvature-sensitive effective decoherence rate $\gamma_{\rm eff}=\gamma_0+\lambda f(R,\bar{B})$, where $f(R,\bar{B})$ is the minimal bilinear form $f(R,B)=R\bar{B}$.
Hence, 
\begin{equation}
\gamma_{\rm eff} = \gamma_0 + \lambda R \bar{B}.
\label{gamma}
\end{equation}

For the initial singlet state $\ket{\psi^-} = \frac{1}{\sqrt2}\big(\ket{01}-\ket{10}\big)$ of two qubits written is the computational basis and the Werner channel \cite{Manzano}, the evolution of the mixed state $\rho(t)$ is given by
\begin{equation}
\rho(t) = p(t)\,\ket{\psi^-}\!\bra{\psi^-} + \frac{1-p(t)}{4}\,I_4,
\end{equation}
where $p(t) = e^{-\gamma_{\rm eff}t}$. The CHSH inequality is violated whenever $2\sqrt2\,p(t) > 2$, i.e.\ $p(t) > 1/\sqrt2$. Writing $t_B$ for the timescale of
the Bell test, this condition reads $e^{-\gamma_{\rm eff}t_B} > 1/\sqrt2$, leading to the upper bound on the coupling constant $\lambda$. 

\begin{equation}
\lambda < \frac{\tfrac12\ln2-\gamma_0 t_B}{R\,\kappa\,t_B}.
\end{equation}
where we used $C(t=0)=1$ and $\bar{B}_0=\kappa$.

It is time to justify using the standard master equation and Eq.~(\ref{gamma}). To do that, let us obtain an effective field equation from the action (\ref{action}). Varying the action with respect to $g^{\mu\nu}$, and holding $h^{ab}$ fixed gives us
\begin{equation}
\left(\frac{1}{16\pi G}+\lambda\bar B\right)\Big(R_{\mu\nu}-\tfrac12 g_{\mu\nu}R\Big) +
\frac{\Lambda}{16\pi G}g_{\mu\nu} = 0,
\end{equation}
thus $\lambda\bar{B}$ may be interpreted as an effective Newton constant, i.e. 
\begin{equation}
\frac{1}{G_{\rm eff}} = \frac{1}{G} + 16\pi\lambda\bar{B},
\end{equation}
meaning that quantum entanglement acts as an effective curvature source. Note that entanglement does not appear as a term on the right-hand side of the Einstein equation the way ordinary stress-energy would \cite{Neukart}.

However, the variation of the action with respect to $h^{ab}$ alone does not change the field equation due to the fact that the interaction term (\ref{interaction}) contains $\bar{B}$, which is an average over the Hilbert manifold. Due to this asymmetry in the definition of the interaction, we can use the relationship (\ref{gamma}), without curvature-dependent modifications of quantum dynamics. 

In conclusion, I would like to stress that the model is intentionally minimal. Its purpose is to show how quantum entanglement (or quantum resources) can be treated as an effective curvature source in an operational setting and how that hypothesis can be constrained through the Bell experiment. Future work should derive the interaction term from a more explicit dynamical framework, extend the resource functional beyond pure two-qubit states, and test the robustness of the bound under alternative noise channels and effective decoherence rate conventions.

\section*{Acknowledgments}
I thank Eugene Demler for the opportunity to visit his group at ETH Zurich.

\section*{Declaration of competing interest}
The authors declare no conflict of interest.

\end{document}